\documentclass[conference, 10pt]{IEEEtran}
\IEEEoverridecommandlockouts
\usepackage{cite}
\usepackage{amsmath,amssymb,amsfonts}

\usepackage{graphicx}
\usepackage{textcomp}
\usepackage{xcolor}

\usepackage{bbm}
\usepackage{url}
\usepackage{subfigure}
\usepackage{pifont}
\usepackage{hyperref}
\usepackage{setspace}
\usepackage{multirow}
\usepackage{cases}
\usepackage{amsthm}
\usepackage{times}
\usepackage{stfloats} % enable figure across two columns floating at the end of one page
\usepackage{booktabs}
\usepackage[T1]{fontenc}
\usepackage{comment}
\usepackage{multicol}

\def\BibTeX{{\rm B\kern-.05em{\sc i\kern-.025em b}\kern-.08em
    T\kern-.1667em\lower.7ex\hbox{E}\kern-.125emX}}

\definecolor{orange}{rgb}{1, .36, .08}

\newcommand{\fakeparagraphnospace}[1]{\noindent\textbf{#1.}}

\usepackage[all]{background}
\usepackage{stackengine}
\setstackEOL{\\}
\setstackgap{L}{\normalbaselineskip}
\SetBgContents{\color{blue}{\tiny \Longstack{PREPRINT - accepted at the International Symposium on Quality Electronic Design (ISQED), 2024.}}}% Set contents
\SetBgPosition{4.8cm,1cm}% Select location
\SetBgOpacity{1.0}% Select opacity
\SetBgAngle{0}% Select rotation of logo
\SetBgScale{1.8}% Select scale factor of logo

\renewcommand{\baselinestretch}{0.984} % This is for the camera-ready
\usepackage{algorithmic,algorithm}
\begin{document}

\title{DECOR: Enhancing Logic Locking Against Machine Learning-Based Attacks
}

\author{
Yinghua Hu,$^1$ Kaixin Yang,$^2$ Subhajit Dutta Chowdhury,$^2$ and Pierluigi Nuzzo$^2$\\
\small 
$^1$Synopsys, Inc., Sunnyvale, CA, USA\\
$^2$Department of Electrical and Computer Engineering, University of Southern California, Los Angeles, CA, USA \\ yinghuah@synopsys.com, \{kaixinya, duttacho, nuzzo\}@usc.edu\\[0ex]
% % To make more spaces just add \\!
% % \\
% % \\
% % \\
}

\maketitle

\begin{abstract}
Logic locking (LL) has gained attention as a promising intellectual property protection measure for integrated circuits. However, recent attacks, facilitated by machine learning (ML), have shown the potential to predict the correct key in multiple LL schemes by exploiting the correlation of the correct key value with the circuit structure. This paper presents a generic LL enhancement method based on a randomized algorithm that can significantly decrease the correlation between locked circuit netlist and correct key values in an LL scheme. Numerical results show that the proposed method can efficiently degrade the accuracy of state-of-the-art ML-based attacks down to around $50\%$, resulting in negligible advantage versus random guessing. 
\end{abstract}


\section{Introduction}\label{sec:intro}
The ever-increasing cost of designing and manufacturing modern integrated circuits (ICs) has boosted a global, decentralized supply chain, often relying on 
third-party foundries for manufacturing~\cite{tan2020benchmarking,rajarathnam2020regds}. Equipped with advanced IC reverse-engineering techniques~\cite{meade2016netlist,rajarathnam2020regds,geist2020relic,subhajitreverse,zhang2022trilock}, rogue agents 
may be able to reconstruct and counterfeit the design, bringing 
substantial financial consequences to IC design companies. 
Furthermore, counterfeit chips usually lack sufficient testing before commercialization and may experience early failures, possibly leading to life-threatening risks in safety-critical applications.

Over the past decade, \emph{logic locking} (LL)~\cite{roy2010ending,rajendran2013fault,yasin2016improving,yasin2016sarlock,hu2020sanscrypt,chowdhury2021enhancing} has gained significant attention as a promising, low-cost countermeasure against IC reverse engineering. 
LL protects a circuit by modifying its original netlist at the logic level and by adding a new set of input ports, called key ports. The locked circuit will then display a different function from the original one. To ``unlock'' the circuit and access its hidden function, a user must provide the correct key, i.e., a specific binary pattern, at the key ports. Even if a malicious agent successfully recovers the locked circuit structure from the design layout, its original function will still be hidden without the correct key, which helps prevent IC counterfeiting. 
\begin{figure}[t]
	\centering
	\subfigure[]{\includegraphics[width=0.48\columnwidth]{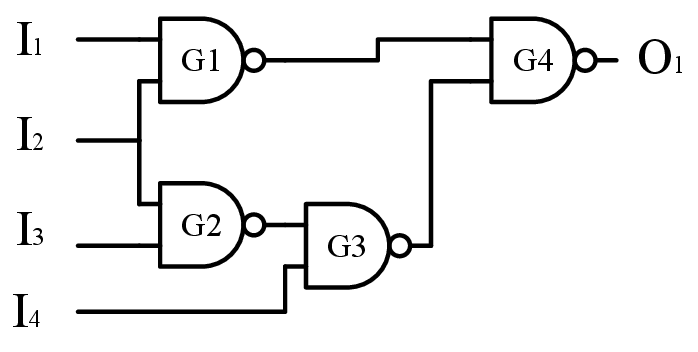}\label{fig:original_example}}
	\subfigure[]{\includegraphics[width=0.48\columnwidth]{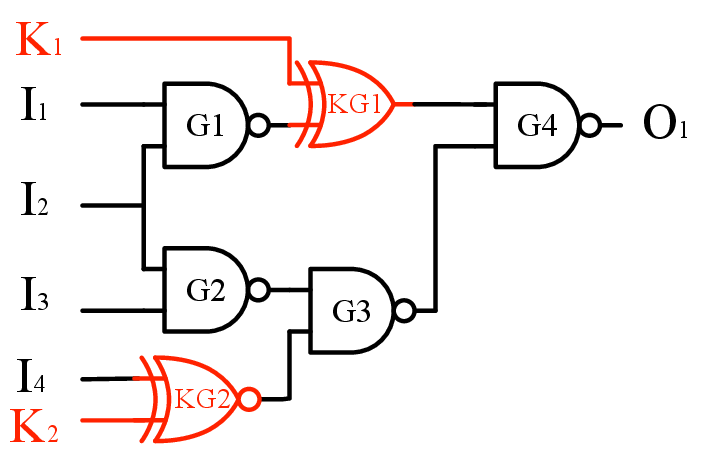}\label{fig:locked_example}}
    \vspace{-3mm}
	\caption{The schematics of a simple circuit (a) before and (b) after being locked by XOR/XNOR-based insertion~\cite{roy2010ending,rajendran2013fault,yasin2016improving}.}
	\vspace{-5mm}
	\label{fig:ll_example}
\end{figure}
For example, the simple circuit in Fig.~\ref{fig:original_example} is locked, as shown in Fig.~\ref{fig:locked_example}, by using an LL technique based on the insertion of XOR/XNOR gates ~\cite{roy2010ending,rajendran2013fault,yasin2016improving}. 
To restore the original circuit functionality, a user must provide ``$0$'' and ``$1$'' to key ports $K_1$ and $K_2$, respectively, so that the gates in red 
act as buffers and do not toggle the signals propagating through them. 

Unfortunately, several LL techniques have been reported as being potentially vulnerable, due to the fact that  
% it has been revealed that 
the netlist structure of a locked circuit may leak critical information about the correct key~\cite{chakraborty2018sail,hu2021risk,sisejkovic2021logic}. 
% Consequently, 
This vulnerability has shown to be exploitable by a family of attacks leveraging various machine learning (ML) models, both non-graph-based~\cite{chakraborty2018sail,sisejkovic2020challenging,raj2023deepattack} and graph-based~\cite{alrahis2021omla}, to identify correlations between the circuit structure and the key, and predict the correct key pattern with high accuracy. 
% by exploiting the strong correlation between the netlist structure and the correct key. 
ML-based attacks are expected to become even more powerful due to the continuous advancements in ML algorithms and computing power, calling for further investigation of the information leakage in locked designs as well as methods that help efficiently decorrelate the locked circuit structure from the correct key.  

Previous efforts to robustify LL against ML-based attacks have mostly focused on making a set of \emph{localized, pre-determined transformations of the circuit structure} with the goal of reducing the correlation between the circuit structure and the correct key~\cite{alrahis2021unsail}. 
However, such approaches do not alter the function of the locked circuit and, thus, may be prone to ``re-synthesis'' attacks~\cite{sisejkovic2021deceptive}.
To prevent this drawback, this paper proposes
\texttt{DECOR}, a randomized algorithm-based construction which involves \emph{strategic modifications to the function of the locked circuit}. 
Since the netlist structure is synthesized from the modified circuit function, \texttt{DECOR} can 
significantly decrease the correlation between the netlist structure and the correct key, 
reducing the expected advantage of an ML-based attack to be   
negligible when compared with a random key guessing attack. 
The flow of \texttt{DECOR} involves only behavioral changes to the locked circuit and is, therefore, applicable to enhance any LL technique. 
Our contributions can be summarized as follows:
\begin{itemize}
  \item We present a method to strategically alter the functionality of a locked circuit in a randomized manner without preventing a legal user from accessing the original circuit function. The modified locked circuits, when used by an ML-based attack, can lead to the creation of misleading training data sets. 
  \item Based on our proposed method, we introduce \texttt{DECOR}, an LL enhancement strategy
  to achieve decorrelation between the netlist structure and the correct key. 
  To the best of our knowledge, \texttt{DECOR} is the first LL enhancement method that can be efficiently applied to any LL scheme to improve its resilience to ML-based attacks exploiting the correlation between circuit structure and key. 
  \item We evaluate the effectiveness of our approach against \texttt{OMLA}~\cite{alrahis2021omla}, an open-sourced, state-of-the-art ML-based attack based on graph neural networks (GNNs), 
  showing its ability to reduce the key prediction accuracy to around $50\%$, rendering \texttt{OMLA}'s advantage negligible over random guessing.
\end{itemize}

The rest of the paper is organized as follows. Section~\ref{sec:background} discusses the threat model underlying the ML-based attacks considered in this paper and some related approaches proposed 
in the literature to counteract them. In Section~\ref{sec:method}, we illustrate the mechanism of \texttt{DECOR} and show how an LL scheme can be efficiently enhanced by \texttt{DECOR} to achieve resilience to ML-based attacks. 
We validate the effectiveness of our approach 
in Section~\ref{sec:exp}. Conclusions are drawn in Section~\ref{sec:conclusion}.

\section{Background and Related Work}\label{sec:background}

\begin{figure}[t]
	\centering
	\includegraphics[width=\columnwidth]{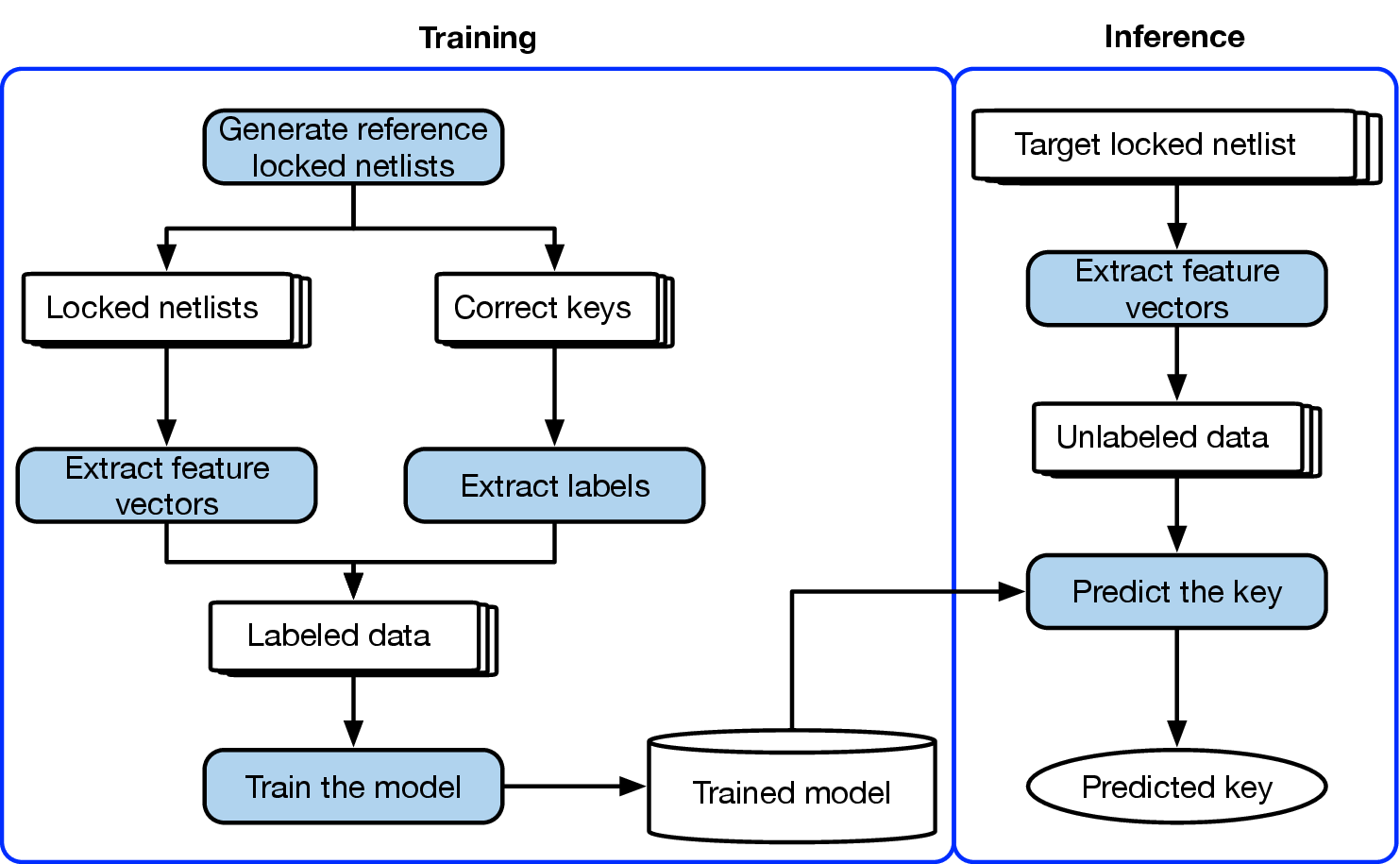}
    \vspace{-5mm}
	\caption{Flowchart of an ML-based attack. 
	}
	\vspace{-5mm}
	\label{fig:ml_attack_flow}
\end{figure}

\begin{figure*}[bp]
	\centering
    \vspace{-6mm}
	\includegraphics[width=0.95\textwidth]{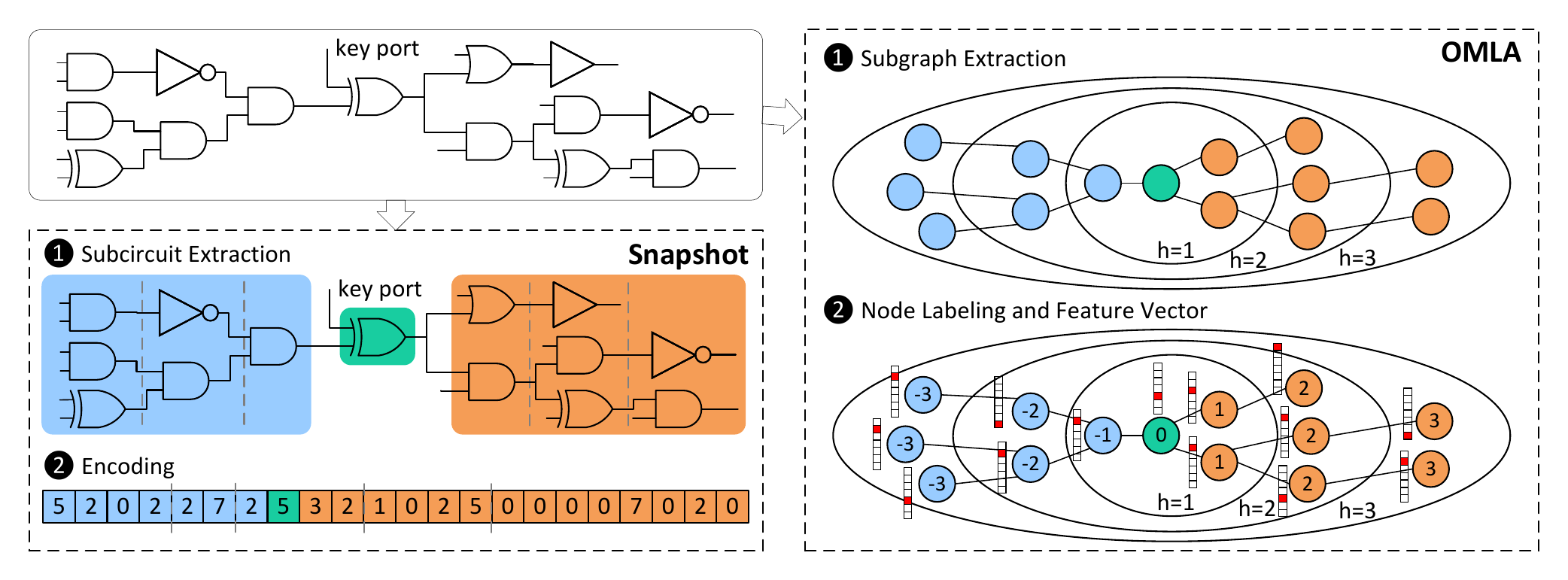}
	\vspace{-3mm}
	\caption{An example of feature vector extraction and encoding in \texttt{SnapShot}~\cite{sisejkovic2020challenging} and \texttt{OMLA}~\cite{alrahis2021omla}.
	}
	\vspace{-8mm}
	\label{fig:locality}
\end{figure*}

We start by illustrating
the threat model used by ML-based attacks and discuss several previous efforts that aim to counteract them. 

\subsection{ML-Based Attacks: Threat Model}\label{subsec:threat_model}

ML-based attacks, e.g., \texttt{SAIL}~\cite{chakraborty2018sail}, \texttt{SnapShot}~\cite{sisejkovic2020challenging}, aim to correctly predict each bit of the correct key of a locked netlist, called a \emph{target netlist}. 
They typically assume that attackers have access to two resources, namely, (i) the target netlist and (ii) the knowledge of the specific LL technique used to lock the target netlist. 
As shown in Fig.~\ref{fig:ml_attack_flow}, an ML-based attack consists of two phases, namely, \emph{training} and \emph{inference}. 
The training phase includes three steps, that is, (i) reference netlist generation, (ii) training data extraction, and (iii) model training. We detail each step below. 

\subsubsection{Reference Netlist Generation}
By using the same LL technique applied to the target netlist, an attacker is able to generate a significant number of locked netlists, called \emph{reference netlists}, whose correct keys are known. 
When generating reference netlists, the attacker may choose to lock a generic benchmark suite, i.e., any set of netlists not including the target one~\cite{sisejkovic2020challenging}, an option often referred to as the \emph{generalized set scenario} (GSS).
Alternatively, they may consider the target netlist as a benchmark and re-lock it to generate reference netlists. The latter option, also called \emph{self-referencing scenario} (SRS), is more commonly used in practice, since it usually achieves better key prediction accuracy~\cite{chakraborty2018sail,sisejkovic2020challenging,alrahis2021omla,raj2023deepattack}. 

\subsubsection{Training Data Extraction}

An important step before training the ML model is to extract, from the reference locked netlists, the training data set that includes both the features (netlist properties) and the labels (correct keys). 
For instance, \texttt{SnapShot}~\cite{sisejkovic2020challenging} associates each feature-label pair with a single key port.
The feature for the key port is a fixed-length vector that captures information about the function and topology of the neighborhood of the key port. 

A specific example illustrating the feature encoding of \texttt{SnapShot} is shown in the left side of Fig.~\ref{fig:locality}, where the value and the position of each entry in the feature vector represent the type and the location, respectively, of each gate in the key port neighborhood. 
To represent the topology of the key port neighborhood with less information loss, another ML-based attack, called \texttt{OMLA}, directly extracts as a feature a subgraph whose topology is identical to that of the key port neighborhood, as shown on the right side of Fig.~\ref{fig:locality}. 
The label associated with the feature is the correct key bit of the corresponding key port, which is either $0$ or $1$. 
All the feature-label pairs extracted from the reference locked netlists constitute the training data set. 

% \fakeparagraphnospace{Model Training} 
\subsubsection{Model Training} 

An ML model is trained with the training data set. % To the best of our knowledge, there is no general criterion for choosing the ML models. 
If the features in the data set are in vector form,
previous attacks have adopted different ML models, such as random forests (RFs)~\cite{chakraborty2018sail}, multi-layer perceptrons (MLPs)~\cite{sisejkovic2020challenging}, convolutional neural networks (CNNs)~\cite{sisejkovic2020challenging}, and deep neural networks (DNNs)~\cite{raj2023deepattack}. 
% for vector-form features, 
With features in the form of a graph, a graph neural network (GNN) has shown to be effective and has been adopted to construct the \texttt{OMLA} attack~\cite{alrahis2021omla}.

After the ML model is trained, in the inference phase, the attacker extracts features from the target netlist with the same encoding adopted in the training phase, and uses them to predict the corresponding key bits. The success of an ML-based attack can be measured by the \emph{key prediction accuracy} (KPA), which is the percentage of correctly predicted key bits among all the key bits in the target locked netlist~\cite{chakraborty2018sail,sisejkovic2020challenging,alrahis2021omla,raj2023deepattack}. 

\begin{table}[t]
    \caption{Comparison of different countermeasures to ML-based attacks.} 
    \vspace{-5mm}
    \begin{center}
    \resizebox{\columnwidth}{!}{
    \begin{tabular}{|c|cccc|c|}
			\hline
			\multirow{2}{*}{\textbf{Countermeasure}} & \multicolumn{4}{c|}{\textbf{Secure against}} & \multirow{2}{*}{\textbf{Applicable to all LL schemes}} \\ \cline{2-5}
			 & \multicolumn{1}{c|}{Re-synthesis attack} & \multicolumn{1}{c|}{ \texttt{SAIL}} & \multicolumn{1}{c|}{ \texttt{SnapShot}} &  \texttt{OMLA} & \\ \hline
			MUX-based~\cite{sisejkovic2021deceptive,chowdhury2023simll} & \multicolumn{1}{c|}{\ding{52}} & \multicolumn{1}{c|}{\ding{52}}  & \multicolumn{1}{c|}{\ding{52}} & \ding{52}  & \color{red}{\ding{54}} \\ \hline
			\texttt{UNSAIL}~\cite{alrahis2021unsail} & \multicolumn{1}{c|}{\color{red}{\ding{54}}} & \multicolumn{1}{c|}{\ding{52}}  & \multicolumn{1}{c|}{\color{red}{\ding{54}}} & \color{red}{\ding{54}} & \color{red}{\ding{54}} \\ \hline
			\texttt{TRLL}~\cite{limaye2020thwarting} & \multicolumn{1}{c|}{\ding{52}} & \multicolumn{1}{c|}{\ding{52}}  & \multicolumn{1}{c|}{\ding{52}} & \ding{52} & \color{red}{\ding{54}} \\ \hline
			\texttt{DECOR} (this paper) & \multicolumn{1}{c|}{\ding{52}} & \multicolumn{1}{c|}{\ding{52}}  & \multicolumn{1}{c|}{\ding{52}} & \ding{52} & \ding{52} \\ \hline
		\end{tabular}
    }
    \label{tab:compare_methods}
    \end{center}
    \vspace{-7mm}
\end{table}

Several LL schemes have shown to induce strong correlations between the structure of the locked circuit and the correct key, making themselves vulnerable to ML-based attacks. 
For example, existing ML-based attacks have all reported a high KPA for XOR/XNOR-based insertion (\texttt{XBI}) locking~\cite{roy2010ending,rajendran2013fault,yasin2016improving}. 
A simple example of \texttt{XBI}-locked netlist is shown in 
% For example, as suggested in 
Fig.~\ref{fig:locked_example}, where the type of the inserted gates, i.e., XOR or XNOR, may directly indicate the correct key value of $0$ or $1$, respectively. 
Logic synthesis may be able to perturb the netlist structure to some extent. However, existing ML-based attacks are still successful in identifying strong correlations in the post-synthesis netlists~\cite{chakraborty2018sail}. 
Another category of LL schemes, including, e.g., \texttt{SARLock}~\cite{yasin2016sarlock} and \texttt{Anti-SAT}~\cite{xie2018anti}, is based on the insertion of point functions, which are known for their effectiveness in thwarting the Boolean satisfiability-based (SAT) attack~\cite{subramanyan2015evaluating}. Although 
% existing 
ML-based attacks have not yet been mounted against 
% to attack 
such schemes, they may still exhibit similar vulnerabilities as in \texttt{XBI}, in that they tend to  
% those schemes are vulnerable due to the reliance of the aforementioned XOR/XNOR property to 
hard-code information about the correct key in the structure of the locked circuit, hence the strong correlation between circuit structure and key~\cite{sisejkovic2020challenging,hu2021risk}.

\subsection{Countermeasures to ML-Based Attacks}

Various methods have been proposed to make LL more resilient to ML-based attacks~\cite{sisejkovic2021logic}. 
\emph{Deceptive Multiplexer} (\texttt{D-MUX})~\cite{sisejkovic2021deceptive} and \texttt{SimLL}~\cite{chowdhury2023simll} propose to lock the circuit by always inserting the same structure, i.e., multiplexers, instead of the distinguishable XOR and XNOR gates, such that the inserted structure is no longer correlated with the correct key. 
Although these locking methods can circumvent ML-based attacks, multiplexer-based insertion is still deemed vulnerable to other attack vectors, such as the SAT attack~\cite{subramanyan2015evaluating}.

\texttt{UNSAIL}~\cite{alrahis2021unsail} aims to circumvent \texttt{SAIL}~\cite{chakraborty2018sail} by replacing repetitive structural patterns in the \texttt{XBI}-locked netlist with a predefined set of alternative patterns. Although \texttt{UNSAIL} can mislead the ML model used in \texttt{SAIL} and lower down the KPA, the predefined structural transformation to the netlists do not involve any functional alteration and can be potentially reversed by a ``re-synthesis'' attack~\cite{sisejkovic2021deceptive}. 

\emph{Truly Random Logic Locking} (\texttt{TRLL})~\cite{limaye2020thwarting} overcomes the above shortcoming of \texttt{UNSAIL} by involving randomness in the key gate selection and implementation. 
However, \texttt{TRLL} is specifically designed for the \texttt{XBI} scheme. 
Unlike previous efforts, which mostly focus on specific attacks or LL schemes, we propose \texttt{DECOR}, an LL enhancement strategy
that is not only applicable to a particular LL technique but  
\emph{can be used to robustify any existing LL scheme
against ML-based attacks}, by reducing the correlation between netlist structure and correct keys.
We summarize the properties of \texttt{DECOR} and the previous countermeasures against ML-based attacks in Table~\ref{tab:compare_methods}.

Besides attempting at predicting the correct key, machine learning has also been applied to detect other aspects of a locked circuit that facilitate reverse engineering.  
For example, \texttt{GNNUnlock}~\cite{alrahis2021gnnunlock} is a recent attack that leverages GNNs to predict whether a gate or module in the locked netlist belongs to a specific block inserted by a particular LL scheme, called \texttt{SFLL}~\cite{yasin2017provably}, and can be removed. 
In this paper, we only focus on ML-based attacks that exploit the correlation between the circuit structure and the key. 
Extensions of the proposed LL enhancement strategy to address correlations
between circuit structures and sensitive information other than the key will be subject of future work.

\section{The \texttt{DECOR} Method}\label{sec:method}

We first discuss how the correlation between the locked circuit structure and the correct key can be reduced via a set of modifications to the function of the locked circuit without affecting the correct operation of the circuit.
Inspired by those functional modifications, we illustrate \texttt{DECOR}, which can be applied to enhance the resilience of an LL method against ML-based attacks. 

\subsection{Preliminaries}

For a circuit $C_o$ with sets $I$ and $O$ of primary input ports and primary output ports, respectively, we denote by $f_o:\mathbb{B}^{|I|}\rightarrow \mathbb{B}^{|O|}$ the circuit function. Let  $K$ be the set of key ports. We denote by $f_l:\mathbb{B}^{|I|}\times \mathbb{B}^{|K|}\rightarrow \mathbb{B}^{|O|}$ 
the function implemented by the locked circuit $C_l$. 
We use $i$ and $k$ to represent the patterns applied at the primary input ports and the key ports, respectively. 

We say that key $k$ is correct if and only if the following condition holds: 
\begin{equation}
    \begin{aligned}
    \forall i \in \mathbb{B}^{|I|}, f_o(i) = f_l(i,k),
    \end{aligned}
\end{equation}
which is abbreviated as $f_o \equiv f_{l,k}$. 
Any locked circuit must have at least one correct key. 

\subsection{
Randomized Alteration of the Locked Circuit Function
}\label{sec:theory}

As discussed in Section~\ref{sec:background}, previous approaches to robustify LL against ML-based attacks have mostly focused on a set of localized, pre-determined transformations of the circuit structure, which do not alter the function of the locked circuit and are vulnerable to ``re-synthesis'' attacks. 
We propose, instead, to achieve the decorrelation between the circuit structure and the correct key by \emph{altering the locked circuit function} according to a 
\emph{randomized} procedure. 

Without loss of generality, we illustrate our approach by assuming that the circuit has only one output port. 
Based on Shannon's expansion theorem, the function of a locked circuit $C_l$ can be decomposed into the disjunction of several Shannon cofactors with respect to the input $k$ as follows,  
\begin{equation}\label{eq:shannon}
    \begin{aligned}
    f_l(i,k)=\bigvee_{k_c\in \mathcal{K}_c}f_l(i,k_c)\vee \bigvee_{k_{w}\in \mathcal{K}_w}f_l(i,k_{w}),
    \end{aligned}
\end{equation}
where $\mathcal{K}_c$ and $\mathcal{K}_w$ are the sets of correct keys and wrong keys, respectively. 
In the special case where only one correct key exists, \eqref{eq:shannon} can be simplified as follows,
\begin{equation}\label{eq:shannon_simp}
    \begin{aligned}
    f_l(i,k) = f_l(i,k^*)\vee \bigvee_{k_{w}\in \mathbb{B}^{|K|}\backslash \{k^*\}}f_l(i,k_{w}),
    \end{aligned}
\end{equation}
where $k^*$ is the correct key and $f_o\equiv f_{l,k^*}$. 

Remarkably, the information represented by the function in~\eqref{eq:shannon_simp} is processed in different ways by a legal user and an ML-based attacker.  
For a legal user, since the goal is to access the circuit function with the correct key $k^*$, the only useful Shannon cofactor 
is $f_l(i,k^*)$. 
Any functional changes to the other cofactors in \eqref{eq:shannon_simp} have no impact on the legal user's access to the correct circuit function. 
Therefore, we call these cofactors \emph{user-don't-care} (UDC) cofactors. 
An ML-based attacker, on the other hand, analyzes the locked netlist as a whole. 
Any functional changes in~\eqref{eq:shannon_simp}, including changes in UDC cofactors, 
could potentially affect the structure of the post-synthesis netlist, hence the outcome of an ML-based attack. 
We can then leverage this information asymmetry to strategically alter the UDC cofactors in~\eqref{eq:shannon_simp} and confuse an ML-based attacker without affecting the legal user.
Specifically, we 
contrive two mapping scenarios, that is, (i) a locked circuit function can be unlocked by multiple keys (\emph{one-to-many mapping}) and (ii) multiple locked circuit functions can be unlocked by the same key
(\emph{many-to-one mapping}). Both scenarios tend to break the one-on-one correspondence between a locked circuit's function and its correct key. Since the circuit structure directly derives from the circuit function via logic synthesis, 
breaking this correspondence also helps decorrelate the netlist structure from the correct key, thus increasing the difficulty of an ML-based attack. 

In the following, we show how we modify the UDC cofactors of the locked circuit function to achieve the two mapping scenarios above. Let us focus on the expression in~\eqref{eq:shannon_simp}, which can represent a locked circuit function independently of the LL scheme. 
Suppose that we select two wrong keys, i.e., $k'$ and $k''$, and modify their corresponding UDC cofactors, i.e., $\hat{f_l}(i,k')$ and $\hat{f_l}(i,k'')$, such that both $k'$ and $k''$ can unlock the original circuit function as well as the correct key $k^*$ does, i.e., 
\begin{equation}\label{eq:more_ckeys}
    \begin{aligned}
    f_o\equiv \hat{f}_{l,k^*} \equiv \hat{f}_{l,k'} \equiv \hat{f}_{l,k''}.
    \end{aligned}
\end{equation}
The modified circuit function can then be re-written as follows, 
\begin{equation}\label{eq:sample_mod}
\footnotesize
    \begin{aligned}
    \hat{f_l}(i,k) = \bigvee_{k_{c}\in \{k^*,k',k''\}}\hat{f_l}(i,k_{c})\vee 
    \bigvee_{k_{w}\in \mathbb{B}^{|K|}\backslash \{k^*,k',k''\}}\hat{f_l}(i,k_{w}).
    \end{aligned}
\end{equation}
The functions in \eqref{eq:shannon_simp} and \eqref{eq:sample_mod} are different, but both of them can be unlocked by $k^*$. 
\begin{figure}[t]
	\centering
	\subfigure[]{\includegraphics[trim={0 3.5cm 0 0},clip,width=\columnwidth]{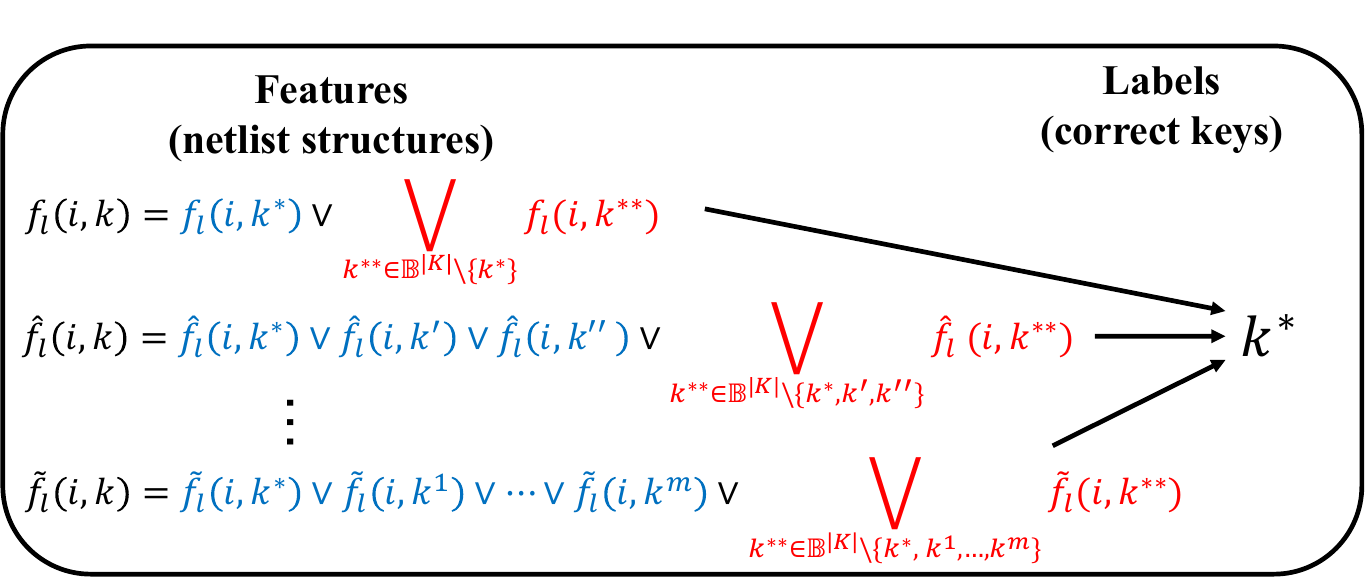}\label{fig:many-to-one}}
	\subfigure[]{\includegraphics[trim={0 2cm 0 0},clip,width=\columnwidth]{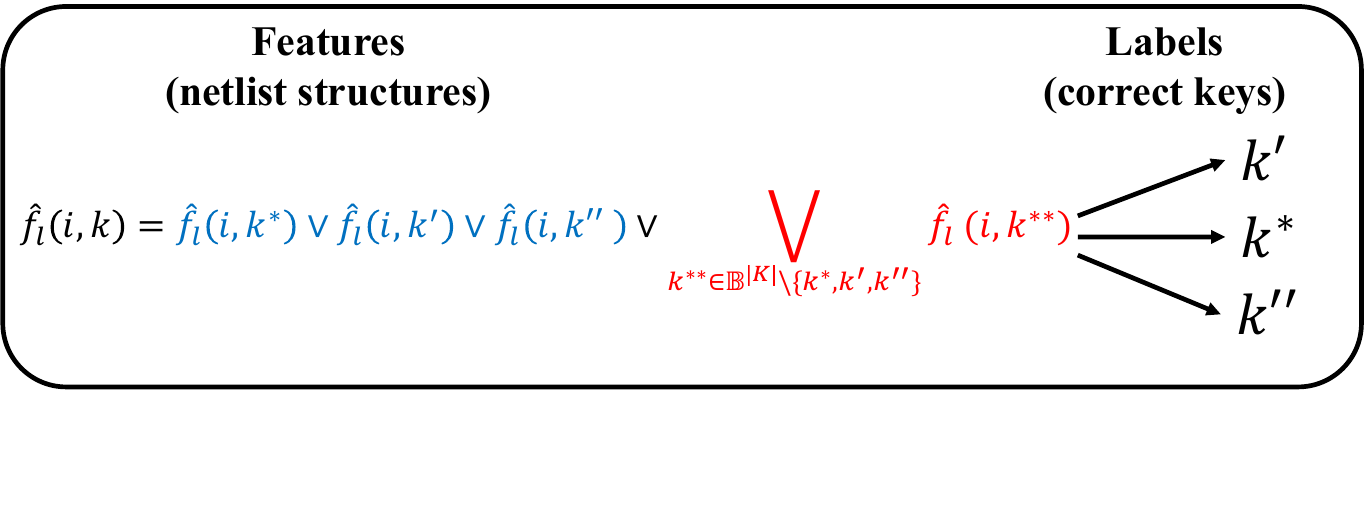}\label{fig:one-to-many}}
    \vspace{-3mm}
	\caption{(a) Many-to-one and (b) one-to-many mappings from features, associated with different locked circuit functions and netlist structures, to correct keys. Each cofactor in blue is equivalent to the original circuit function. Otherwise, the cofactor is in red.}
	\vspace{-3mm}
	\label{fig:mapping}
\end{figure}
In fact, from the expression in \eqref{eq:shannon_simp}, we can generate many different functions for the locked circuit; each of them can be obtained by modifying the UDC cofactors corresponding to one or more arbitrarily selected wrong keys. 
As these circuits are used to generate the training data set for an ML-based attack, the training data set will incorporate the many-to-one mapping scenario, as shown in Fig.~\ref{fig:many-to-one}. 
Symmetrically, these locked circuits also realize the one-to-many mapping scenario. 
For example, as illustrated in Fig.~\ref{fig:one-to-many}, the locked circuit implementing the function in \eqref{eq:sample_mod} 
can be unlocked by any of the three correct keys, i.e., $k^*$, $k'$, or $k''$. Therefore, in the training data set, this circuit can be potentially labeled by three keys.

\subsection{\texttt{DECOR}: Structure-Key Decorrelation Method
}\label{sec:alg}

In Section~\ref{sec:theory}, we demonstrated the possibility of altering a set of UDC cofactors of the locked circuit function
to confuse an ML model. Based on this observation,
we introduce \texttt{DECOR}, a structure-key decorrelation method that uses a randomized algorithm to alter the UDC cofactors of the locked circuit function. In this respect, rather than an LL scheme, \texttt{DECOR} is an enhancement method that can increase the resilience of any LL scheme $\mathcal{L}$ to ML-based attacks.    
For brevity, we denote by \texttt{DECOR}-$\mathcal{L}$ the resulting LL scheme enhanced by \texttt{DECOR}.

Algorithm~\ref{alg:generic_flow} outlines the flow of \texttt{DECOR}-$\mathcal{L}$. 
\begin{algorithm}[t]
 \caption{\texttt{DECOR}-$\mathcal{L}$}
 \begin{algorithmic}[1]\label{alg:generic_flow}
 \renewcommand{\algorithmicrequire}{\textbf{Input:}}
 \renewcommand{\algorithmicensure}{\textbf{Output:}}
\REQUIRE Original circuit $C_{o}$, key size $\kappa$, maximum allowed number of correct keys $N$
\ENSURE  Locked circuit $C_l$, correct key $k^*$

\STATE $f_{int}, k^* = \mathcal{L}(C_{o},\kappa)$

\STATE $correct\_key\_list =[k^*]$

\STATE $n =$ random\_integer\_gen$(range(2,N))$
\WHILE{\textbf{True}}
    \IF{length$(correct\_key\_list) \geq n$}
        \STATE break
    \ENDIF
    \STATE $new\_key = $ random\_key\_gen$(\kappa)$
    \IF{$new\_key {\rm\ not\ in\ } correct\_key\_list$}
        \STATE $correct\_key\_list$.append($new\_key$)
    \ENDIF
\ENDWHILE

\STATE $f_l=$ alter\_UDC\_cofactors$(f_{int},correct\_key\_list)$

\STATE $C_l=$ synthesize$(f_l)$

\RETURN $C_l, k^*$
\end{algorithmic} 
\end{algorithm}
Like any other LL schemes, \texttt{DECOR}-$\mathcal{L}$ takes as input the original circuit $C_o$ and the user-specified key size $\kappa$. In addition, a user must also specify the maximum allowed number of correct keys $N$ for the locked circuit $C_l$. When the algorithm finishes, it returns the locked circuit $C_l$ and one of the correct keys $k^*$. 
\texttt{DECOR}-$\mathcal{L}$ starts by locking the original circuit $C_o$ with the LL scheme $\mathcal{L}$, which generates an intermediate locked circuit function, denoted as $f_{int}$, and a correct key, denoted as $k^*$ (line 1). 
The correct key $k^*$ is then stored in the list of correct keys in line 2. 
Next, an integer $n$ is randomly sampled from the range of $2$ to $N$, which determines the number of correct keys that the locked circuit $C_l$ should have (line 3). 
From line 4 to line 12, the algorithm randomly chooses $n-1$ wrong keys and adds them to the list of correct keys. 
Finally, it modifies the UDC cofactors of $f_{int}$ that correspond to those $n-1$ wrong keys, such that they can also unlock the original circuit function $f_o$ (line 13). 

\begin{figure}[t]
	\centering
	\vspace{-5mm}
	\subfigure[]{\includegraphics[width=0.48\columnwidth]{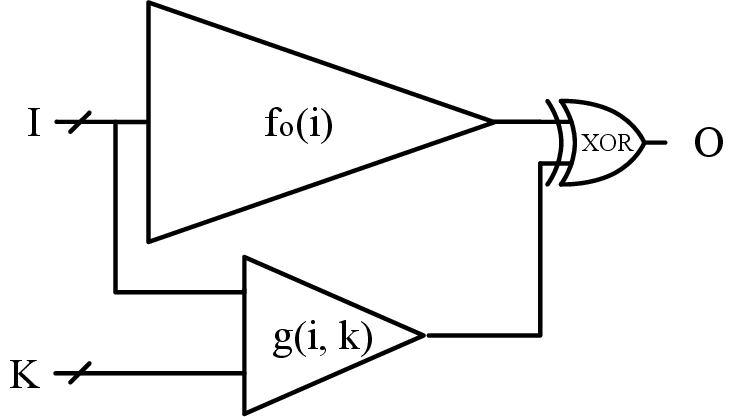}\label{fig:vulnerable}}
	\subfigure[]{\includegraphics[width=0.48\columnwidth]{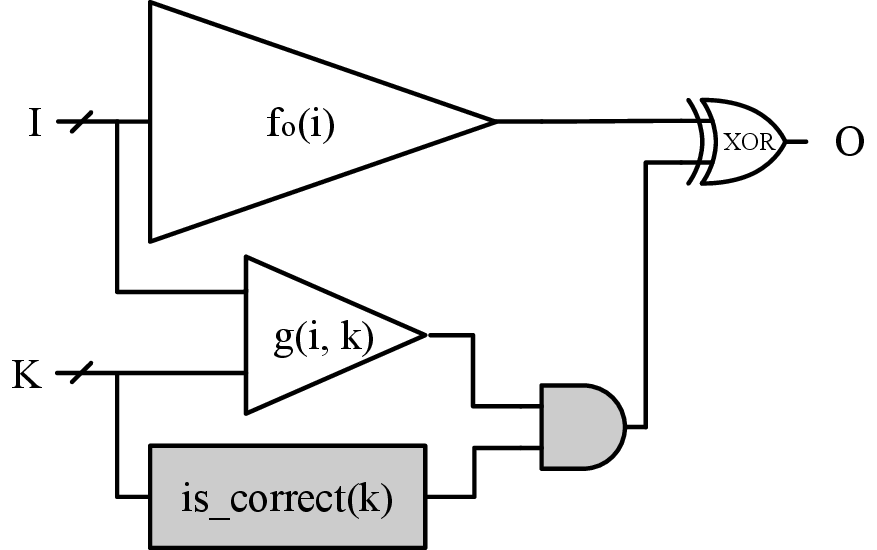}\label{fig:non_vulnerable}}
	\vspace{-3mm}
	\caption{Functional models of (a) the intermediate locked circuit function $f_{int}$~\cite{zhou2017humble,hu2019models} and (b) the locked circuit function $f_{l}$ after adding additional correct keys.}
	\vspace{-5mm}
	\label{fig:add_key}
\end{figure}

The intermediate locked circuit function $f_{int}$, as shown in Fig.~\ref{fig:vulnerable}, can be generally modeled as the exclusive-OR of the original circuit function $f_o$ and a newly introduced function $g$ which represents the effect of $\mathcal{L}$~\cite{zhou2017humble,hu2019models}. 
The effect of altering a set of UDC cofactors can then be obtained by inserting a functional module, called ``is\_correct'', as in Fig.~\ref{fig:non_vulnerable}, which outputs zero for all the keys listed in $correct\_key\_list$ and restores the original circuit function $f_o$ at the circuit output. 
The UDC cofactor alteration is specified at the behavioral level, e.g., via the \textsc{Verilog} representation of the intermediate locked circuit function $f_{int}$. The locked circuit function $f_l$ is then used to generate the netlist $C_l$ via logic synthesis. $C_l$ is returned as output along with the first correct key $k^*$. 
In the following, we show that \texttt{DECOR}-$\mathcal{L}$
can generate  
one-to-many and many-to-one mapping scenarios in the training data set with high probability.
Specifically, we show by contradiction that the following two scenarios are very unlikely to happen. 

\fakeparagraphnospace{(1) If Algorithm~\ref{alg:generic_flow} generates $t$ identical locked circuits, the probability that it reports the same correct key is very small}  
Since the $t$ netlists are identical, 
they must share the same list of correct keys. Suppose the list of correct keys contains $n$ keys.
Since each of the $n$ key values has the same probability of being returned 
% by Algorithm~\ref{alg:generic_flow} 
as the correct key,
% Then, 
the probability that the $t$ runs return the same correct key, hence the same label in the training data set, is 
\begin{equation}\label{eq:prob_1}
    \begin{aligned}
      Pr[{\rm Same\ label\ }| {\rm\ Same\ feature}] = \frac{1}{n^{t-1}}\leq \frac{1}{2^{t-1}},
    \end{aligned}
\end{equation}
where the upper bound is due to $n$ ranging from $2$ to $N$.

\begin{figure*}[t]
	\centering
	\includegraphics[width=\textwidth]{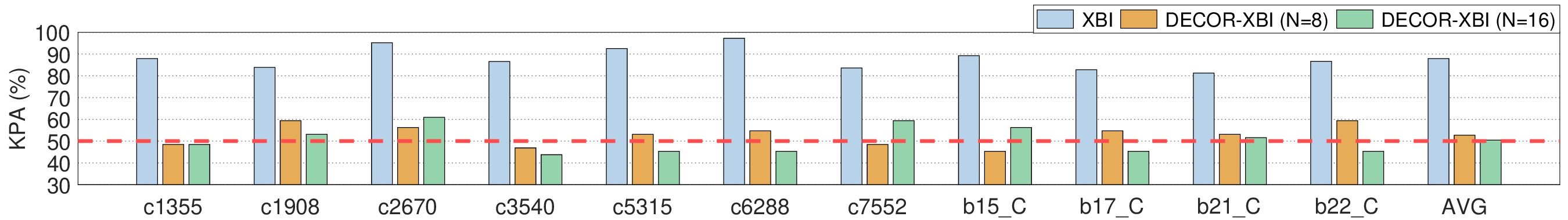}
    \vspace{-6mm}
     \caption{Key prediction accuracy of \texttt{OMLA} on \texttt{XBI} and \texttt{DECOR-XBI}. 
	}
    \vspace{-2mm}
	\label{fig:xbi_first_kpa_result}
\end{figure*}
\begin{figure*}[t]
	\centering
	\includegraphics[width=\textwidth]{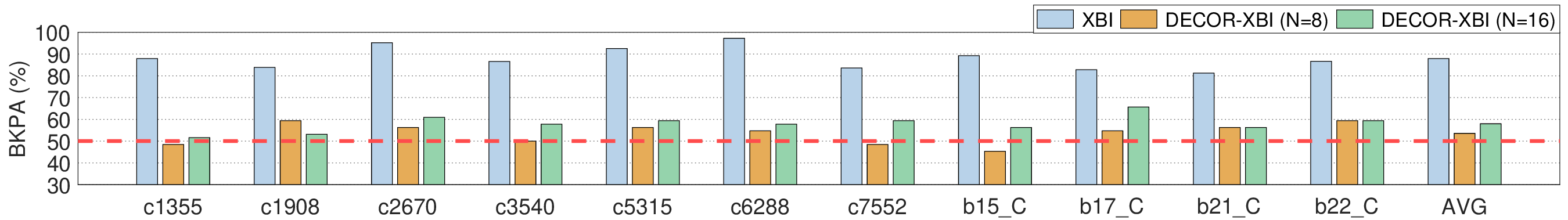}
    \vspace{-6mm}
	\caption{Best key prediction accuracy of \texttt{OMLA} on \texttt{XBI} and \texttt{DECOR-XBI}. 
	}
    \vspace{-2mm}
	\label{fig:xbi_best_kpa_result}
\end{figure*}
\begin{figure*}[t]
	\centering
	\includegraphics[width=\textwidth]{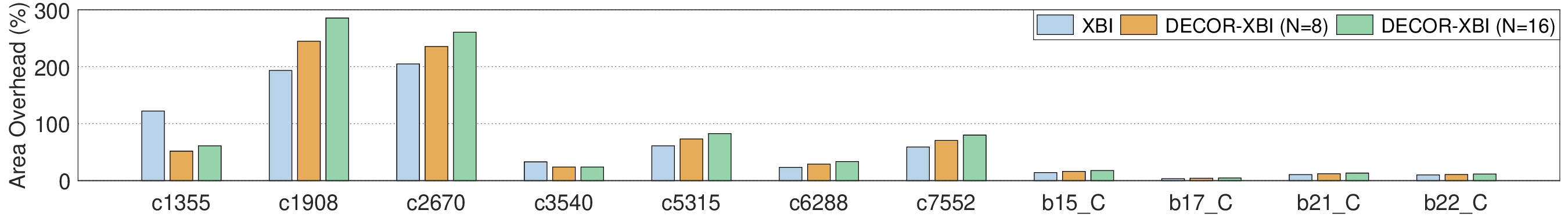}
    \vspace{-6mm}
	\caption{Area overhead for different circuits locked by \texttt{XBI} and \texttt{DECOR-XBI}.}
 \vspace{-3mm}
	\label{fig:xbi_overhead}
\end{figure*}

\fakeparagraphnospace{(2) If Algorithm~\ref{alg:generic_flow} generates $t$ locked circuits with the same reported key, the probability that 
these circuits are identical is very small}
We denote the number of correct keys for the $t$ generated locked circuit functions by
$n_1$, $n_2$, \ldots, $n_t$, respectively. 
For the $t$ functions of the locked circuits to be identical, they must share the same list of correct keys, hence the same number of correct keys, i.e., $n_1=n_2=\cdots=n_t$, and the same key values. 
Given that the number of correct keys is randomly sampled between $2$ and $N$, the probability that the $t$ locked circuit functions have the same number $i$ of correct keys is $1/{(N-1)^{t}}$. 
Besides the key returned by Algorithm~\ref{alg:generic_flow}, the rest of the $i-1$ correct keys should also be the same across the $t$ locked circuit functions. 
Since those $i-1$ correct keys are randomly sampled from the remaining key space of size $2^\kappa-1$, the probability that the rest of the $i-1$ correct keys are the same across the $t$ locked circuit functions
will be $1/\binom{2^{\kappa}-1}{i-1}^{t-1}$.
Finally, the probability of $t$ identical correct keys, i.e., labels, being associated with the same locked circuit function can be expressed 
as follows,
\begin{equation}\label{eq:prob_2}
\small
    \begin{aligned}
    Pr[&{\rm Same\ feature\ }| {\rm\ Same\ label}]\\
    &=\sum_{i=2}^N{Pr[{\rm Same\ feature\ }| {\rm\ Same\ label,\ }n=i]}\cdot Pr[n=i]\\
    &=\sum_{i=2}^N{\frac{1}{\binom{2^\kappa-1}{i-1}^{t-1}}\cdot \frac{1}{(N-1)^{t}}}
    \leq \sum_{i=2}^N{\frac{1}{(2^\kappa-1)^{t-1}} \cdot \frac{1}{(N-1)^{t}}}\\
    &= \frac{(N-1)}{(2^\kappa-1)^{t-1}}\cdot \frac{1}{(N-1)^{t}}\\
    &=\left(\frac{1}{(2^\kappa-1)(N-1)}\right)^{t-1}.
    \end{aligned}
% }
\end{equation}
Both the probabilities in \eqref{eq:prob_1} and \eqref{eq:prob_2} decrease exponentially as $t$ increases. Therefore, when a large set of reference netlists is generated for training the ML model, 
% by our locking flow, 
it is significantly more likely to see different keys labeling the same netlist structure and different circuit structures being labeled by a common key, hence the prevalence of one-to-many and many-to-one mappings in the training data set.

\begin{table}[t]
    \caption{Overview of the selected benchmark circuits} %\revise{To be corrected}}
    \vspace{-6mm}
    \begin{center}
    \resizebox{\columnwidth}{!}{
    \begin{tabular}{|c|c|c|c|c|c|c|c|c|c|c|c|}
    \hline
    \textbf{Circuit} & \textbf{c1355} & \textbf{c1908} & \textbf{c2670} & \textbf{c3540} & \textbf{c5315} & \textbf{c6288} & \textbf{c7552} & \textbf{b15\_C} & \textbf{b17\_C} & \textbf{b21\_C} & \textbf{b22\_C} \\\hline
    \textbf{Inputs} & 41 & 33 & 233 & 50 & 178 & 32 & 207 & 485 & 1452 & 522 & 767 \\\hline
    \textbf{Outputs} & 32 & 25 & 64 & 22 & 123 & 31 & 107 & 449 & 1447 & 509 & 750 \\\hline
    \textbf{Gates} & 455 & 880 & 1193 & 1669 & 2307 & 2661 &  3512 & 8367 & 30777 & 20027 & 29162 \\\hline
    \end{tabular}
    }
    \label{tab:benchmark}
    \end{center}
    \vspace{-4mm}
\end{table}

\section{Validation Results}\label{sec:exp}
We apply \texttt{DECOR} to  
two representative LL schemes, namely, \texttt{XBI}~\cite{roy2010ending} and \texttt{SARLock}~\cite{yasin2016sarlock}, which were conceived to efficiently achieve two major security objectives, i.e., confidentiality, by corrupting the circuit function, and SAT-attack resilience, respectively~\cite{zhou2017humble,hu2019models}. We focus on these schemes, since their underlying mechanisms have been widely adopted in many other LL techniques. 
As discussed in Section~\ref{sec:background}, \texttt{XBI} has already been broken by ML-based attacks\cite{chakraborty2018sail,sisejkovic2020challenging,alrahis2021omla}. 
\texttt{SARLock} is also deemed as being vulnerable~\cite{sisejkovic2020challenging,hu2021risk}, even though a successful ML-based attack has not yet been reported. 
In the following, we validate the effectiveness of \texttt{DECOR}-\texttt{XBI} and \texttt{DECOR}-\texttt{SARLock} against two state-of-the-art ML-based attacks, 
namely, \texttt{OMLA}~\cite{alrahis2021omla} and \texttt{SnapShot}~\cite{sisejkovic2020challenging}, which use graph-based features and vector-based features, respectively. 
Specifically, we provide the experimental results on the full attack flow of \texttt{OMLA}. Since \texttt{SnapShot} is not publicly available, 
% , 
we provide, instead, results from the analysis of the training data set extracted by \texttt{SnapShot}, indicating the good performance of \texttt{DECOR} with non-graph-based attacks.

\begin{figure*}[t]
	\centering
	\includegraphics[width=\textwidth]{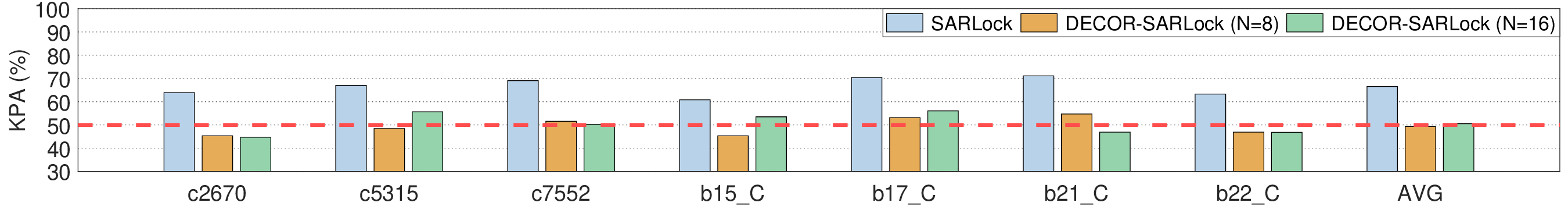}
    \vspace{-6mm}
	\caption{Key prediction accuracy of \texttt{OMLA} on \texttt{SARLock} and \texttt{DECOR-SARLock}. 
	}
    \vspace{-2mm}
	\label{fig:sar_first_kpa_result}
\end{figure*}
\begin{figure*}[t]
	\centering
	\includegraphics[width=\textwidth]{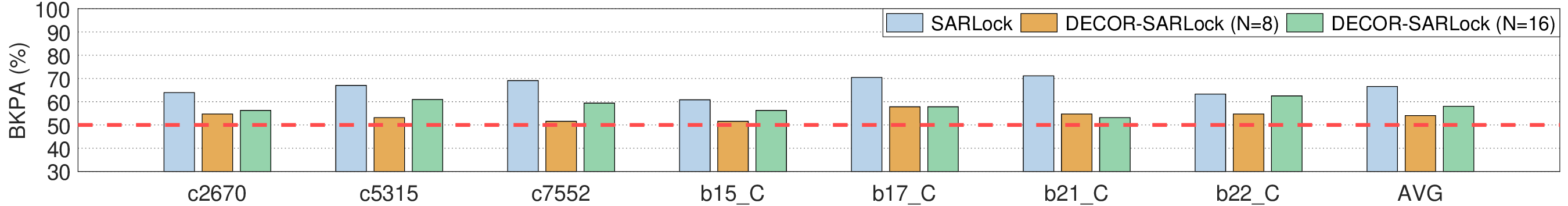}
    \vspace{-6mm}
	\caption{Best key prediction accuracy of \texttt{OMLA} on \texttt{SARLock} and \texttt{DECOR-SARLock}. 
	}
    \vspace{-2mm}
	\label{fig:sar_best_kpa_result}
\end{figure*}
\begin{figure*}[t]
	\centering
	\includegraphics[width=\textwidth]{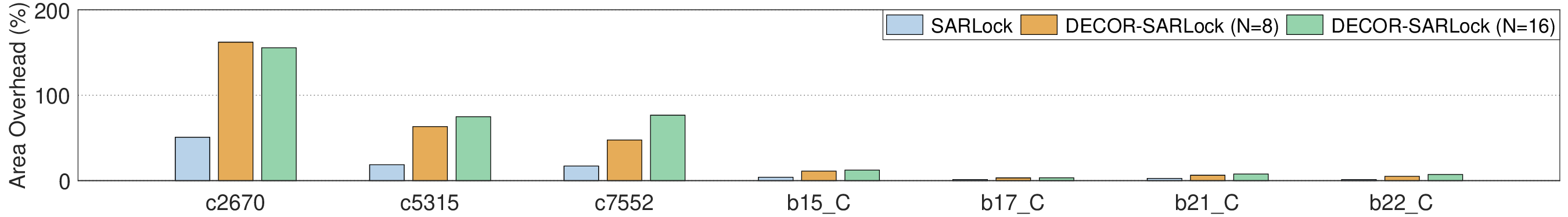}
    \vspace{-6mm}
    \caption{Area overhead for different circuits locked by \texttt{SARLock} and \texttt{DECOR-SARLock}.}
    \vspace{-3mm}
	\label{fig:sarlock_overhead}
\end{figure*}

\subsection{Experiment Setup}

Both \texttt{XBI} and \texttt{SARLock}, as well as their enhanced versions, \texttt{DECOR}-\texttt{XBI} and \texttt{DECOR}-\texttt{SARLock}, are
% The locking method in Algorithm~\ref{alg:generic_flow} is 
implemented in \textsc{Python}. 
We apply these four LL schemes to $11$ different benchmark circuits selected from ISCAS'85 \cite{hansen1999unveiling} and ITC'99 \cite{davidson1999itc}, as detailed in Table~\ref{tab:benchmark}.
For each benchmark circuit, 
% and each LL scheme, 
we generate a target netlist with a key size of $64$. 
Each target netlist is then re-locked with the same LL scheme and the key size of $64$ to generate $1000$ reference netlists used to train the ML model.
In this paper, we adopt $8$ and $16$ as the maximum number of correct keys $N$ to configure Algorithm~\ref{alg:generic_flow}. 
All locked circuits are synthesized with a 45-nm Nangate Open Cell Library~\cite{nangate}. 
The generation of the locked netlists, the synthesis and the ML-based attacks are executed on a server with $48$ $2.1$-GHz cores and $500$-GB memory.
% operated by Ubuntu $18.04.6$. 

Aligned with the literature, we employ the key prediction accuracy (KPA) metric to evaluate the performance of an ML-based attack, indicating the level of protection of \texttt{DECOR}. 
Furthermore, due to the multiple hidden correct keys introduced by \texttt{DECOR}, 
it is necessary to calculate the KPA for each of the correct keys and report the highest KPA among them. In this paper, we call this new metric the \emph{best KPA}, or BKPA in short.

\subsection{XBI vs. DECOR-XBI}

Fig.~\ref{fig:xbi_first_kpa_result} shows the KPA of \texttt{OMLA} on both \texttt{XBI} and \texttt{DECOR-XBI}. Without the enhancement by \texttt{DECOR}, the average KPA across all the benchmark circuits is $87.89\%$, with the KPA of c6288 being as high as $97.22\%$. 
By applying \texttt{DECOR}, the average KPA drops significantly to $52.70\%$ and $50.43\%$, when $N$ = 8 and $N$ = 16, respectively. This demonstrates the ability of \texttt{DECOR} to give ML-based attacks nearly negligible advantage over random guessing, i.e., $50\%$ KPA. 

Similarly, Fig.~\ref{fig:xbi_best_kpa_result} shows the BKPA for the same set of experiments. For \texttt{XBI} without \texttt{DECOR}, the BKPA is equal to the KPA, as there is only one correct key. 
With \texttt{DECOR}, the BKPA can be equal to or larger than the KPA because of the introduction of multiple correct keys. On average, the BKPA for all the benchmark circuits is $53.55\%$ and $57.95\%$, when $N = 8$ and $N = 16$, respectively. 
Despite being slightly higher than the KPA, the BKPA of \texttt{DECOR}-enhanced LL schemes is still much lower than that without applying \texttt{DECOR}. 
% }

Fig.~\ref{fig:xbi_overhead} shows the area overhead of all the target locked netlists. 
For each benchmark circuit, implementing \texttt{DECOR} on the baseline LL scheme, \texttt{XBI}, raises the overhead. 
However, the overhead is inversely proportional to the size of the original circuit. 
% \kaixin{
For example, implementing \texttt{DECOR-XBI} ($N=16$) on the largest benchmark circuit b17\_C incurs only $3\%$ of area overhead. 
The area overhead on small benchmarks is relatively higher. However, when applying \texttt{DECOR} to critical modules in a large system-on-chip, the amortized overhead for the system can still be acceptable.

\subsection{SARLock vs. DECOR-SARLock}

We demonstrate the performance of \texttt{DECOR} on \texttt{SARLock}, another LL scheme that is vulnerable to ML-based attacks. 
Fig.~\ref{fig:sar_first_kpa_result} shows the KPA for each benchmark circuit achieved by \texttt{OMLA}. 
While \texttt{OMLA} achieves an average KPA of $66.52\%$ on the baseline \texttt{SARLock} scheme,  
% Nevertheless, 
\texttt{DECOR-SARLock} can still lower the KPA down to $49.33\%$ and $50.55\%$ when $N=8$ and $N=16$, respectively. 
We observe that four benchmark circuits, namely, c1355, c1908, c3540, and c6288, are excluded from Fig.~\ref{fig:sar_first_kpa_result}. For these four circuits, we observe a KPA of $51.56\%$, $51.56\%$, $53.13\%$, and $51.56\%$ already for the baseline \texttt{SARLock}, 
% the KPA for the baseline \texttt{SARLock} 
which indicates a good resilience of this method to \texttt{OMLA}. This resilience stems from the fact that the key size of the reference netlists used for training the ML model is $128$, which is
larger than the primary input size of the original benchmark circuits. Under this condition, \texttt{SARLock} itself can already generate, as a byproduct, more than one correct key~\cite{yasin2016sarlock,hu2019models} for the same circuit, achieving a similar effect as \texttt{DECOR-SARLock}.  
The BKPA is shown in Fig.~\ref{fig:sar_best_kpa_result}. Compared with the KPA, the average BKPA increases to $54.02\%$ and $58.04\%$, respectively, when $N=8$ and $N=16$. 
The area overhead for all the target netlists locked by \texttt{SARLock} and \texttt{DECOR-SARLock} is shown in Fig.~\ref{fig:sarlock_overhead}, leading to similar observations as for \texttt{DECOR-XBI} in  Fig.~\ref{fig:xbi_overhead}.

\subsection{Analysis of Training Data for SnapShot}

To further illustrate the effectiveness of \texttt{DECOR} against another representative ML-based attack, we analyze the feature vectors used in \texttt{SnapShot}.  
\begin{figure}[t]
	\centering
	% \vspace{-5mm}
	\subfigure[]{\includegraphics[width=0.48\columnwidth]{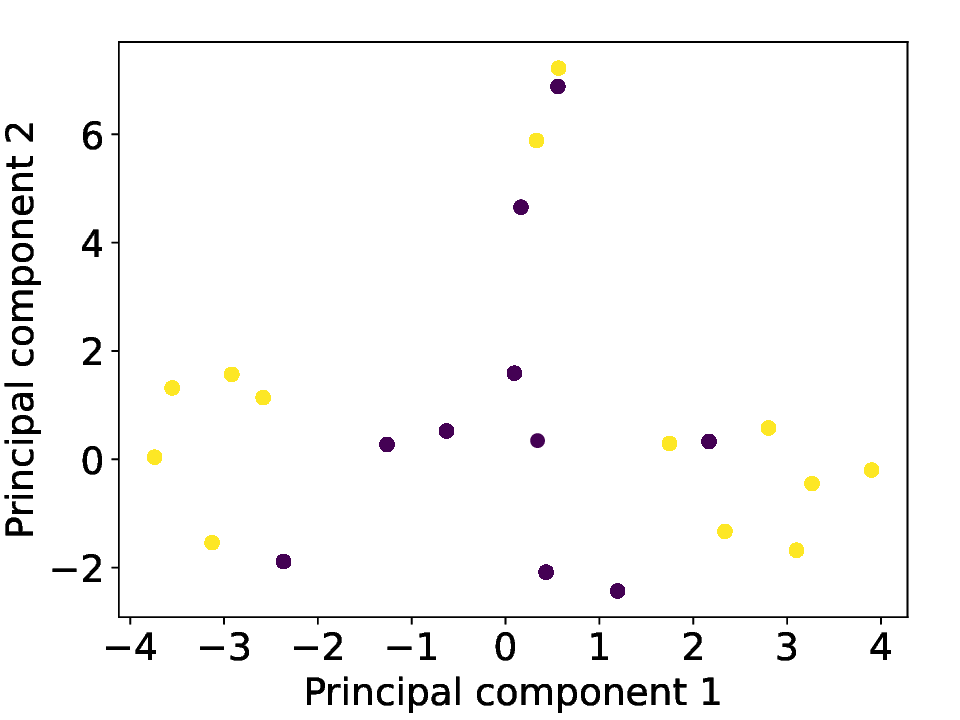}\label{fig:sarlock_pca}}
	\subfigure[]{\includegraphics[width=0.48\columnwidth]{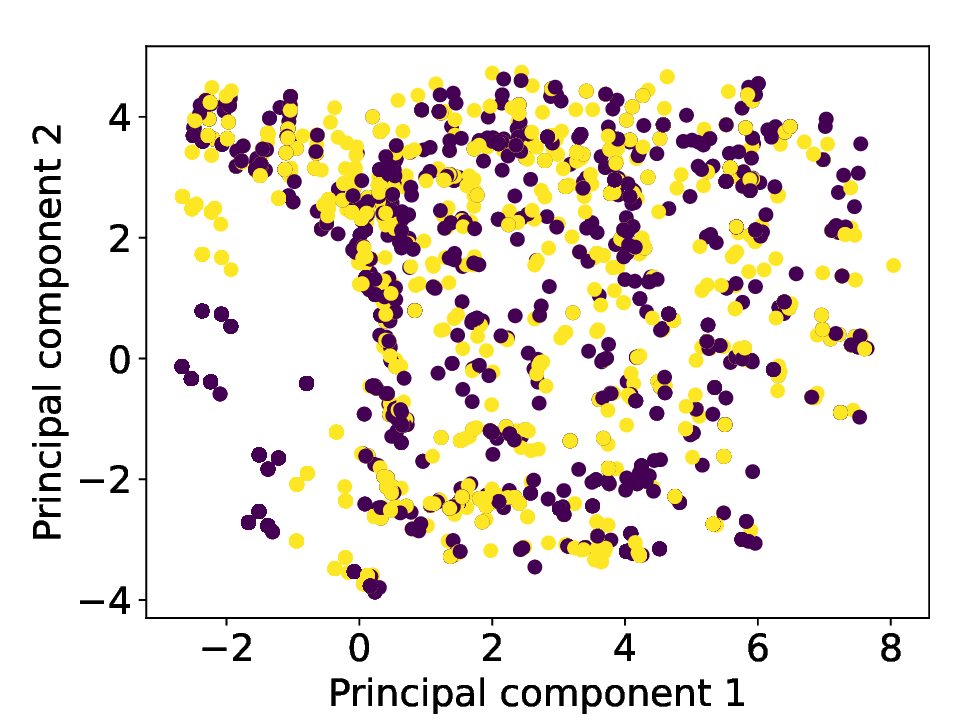}\label{fig:DECOR-sarlock_pca}}
	% \vspace{-3mm}
	\caption{PCA results from the training data set extracted for c2670, which is locked by 
	(a) \texttt{SARLock} and (b) \texttt{DECOR-SARLock}. The two colors denote different labels, i.e., 0 and 1.
	}
	\vspace{-3mm}
	\label{fig:pca}
\end{figure}
Fig.~\ref{fig:pca} shows the results from principal component analysis (PCA) of the $64000$ features extracted from the training data sets for both (a) the baseline \texttt{SARLock} and (b) the improved version \texttt{DECOR-SARLock} ($N=16$). The two colors denote different labels, i.e., correct key bits. 
For \texttt{SARLock}, the purple nodes are clustered around the center of Fig.~\ref{fig:sarlock_pca}, while the yellow ones are distributed away from the center, which indicates a strong correlation between features and labels. On the other hand, in Fig.~\ref{fig:DECOR-sarlock_pca}, the feature distributions for the two labels largely overlap with each other, showing the effectiveness of \texttt{DECOR} in decorrelating features from labels by inserting many-to-one and one-to-many mapped data in the training set.

\section{Conclusions}\label{sec:conclusion}

\sloppypar
We presented \texttt{DECOR}, an efficient and generic structure-key decorrelation method based on randomized circuit function alterations, which can enhance any logic locking scheme to circumvent state-of-the-art ML-based key-prediction attacks. 
Future work includes validating \texttt{DECOR}'s impact on the resilience against other efficient attacks, e.g.,~\cite{li2022redundancy,hu2023security}, and 
extending \texttt{DECOR} to prevent other types of ML-based attacks (see, e.g., \cite{alrahis2021gnnunlock}) which target the prediction of sensitive information that correlates with the locked circuit structure other than the correct key.

\section*{Acknowledgments}
This work was supported in part by the AFRL
and DARPA
under agreement number FA8650-18-1-7817, the NSF
under award 1846524, the Okawa Research Grant, and the USC Center for Autonomy and Artificial Intelligence. 

% ---------Reference---------
\bibliography{reference_list} 
\bibliographystyle{ieeetr}
% ---------------------------
\end{document}